\documentclass[graybox]{svmult}

\usepackage{mathptmx}  
\usepackage{helvet}       
\usepackage{courier}     
\usepackage{type1cm} 
\usepackage{makeidx} 
\usepackage{graphicx}
\usepackage[caption=false]{subfig}
\usepackage{epstopdf}
\usepackage{multicol} 
\usepackage[bottom]{footmisc}
\usepackage{amsmath}
\usepackage{amssymb}
\usepackage{hyperref}

\makeindex

\begin{document}

\title*{The spectrum of gravitational waves in an $f(R)$ model with a bounce.}

\author{Mariam Bouhmadi-L\'{o}pez, Jo\~{a}o Morais and Alfredo B. Henriques}

\institute{Mariam Bouhmadi-L\'{o}pez \at Instituto de Estructura de la Materia, IEM-CSIC, Serrano 121, 28006 Madrid, Spain \email{mariam.bouhmadi@iem.cfmac.csic.es}
\and Jo\~{a}o Morais \at CENTRA, Dept. de F\'{\i}sica, Instituto Superior T\'ecnico, Av. Rovisco Pais 1, 1049 Lisboa, Portugal \email{joao.morais@ist.utl.pt}
\and Alfredo B. Henriques \at CENTRA, Dept. de F\'{\i}sica, Instituto Superior T\'ecnico, Av. Rovisco Pais 1, 1049 Lisboa, Portugal \email{alfredo.henriques@fisica.ist.utl.pt}}

\maketitle


\abstract*{We present an inflationary model preceded by a bounce in a metric $f(R)$ theory. In this model, modified gravity affects only the early stages of the universe. We analyse the predicted spectrum of the gravitational waves in this scenario using the method of the Bogoliubov coefficients. We show that there are distinctive (oscillatory) signals on the spectrum for very low frequencies; i.e., corresponding to modes that are currently entering the horizon.}

\abstract{We present an inflationary model preceded by a bounce in a metric $f(R)$ theory. In this model, modified gravity affects only the early stages of the universe. We analyse the predicted spectrum of the gravitational waves in this scenario using the method of the Bogoliubov coefficients. We show that there are distinctive (oscillatory) signals on the spectrum for very low frequencies; i.e., corresponding to modes that are currently entering the horizon.}


\section{Introduction}
\label{sec:1}

We propose a bouncing scenario \cite{Novello1} within the context of a metric $f(R)$ theory \cite{Capozziello:2011et}:
\begin{equation}
	\label{eq: fR_action}
	S = \frac{1}{2\kappa^2}\int d^4x\sqrt{-g}f(R) + S^{(m)}.
\end{equation}
The bounce we will consider in our model is followed by an inflationary era which is asymptotically de Sitter where, in addition, the gravitational action approaches the Hilbert-Einstein action on that regime, such that the modification to Einstein's General Relativity (GR) affects exclusively the very early universe, around the bounce and a few e-folds after that. We will constrain the model obtaining the spectrum of the stochastic gravitational fossil as would be measured today.

\section{Model for the early universe}
\label{sec:2}
Inspired on the de Sitter solution for a closed  space-time, we define the scale factor around the bounce as:
\begin{equation}
	\label{eq: scalefactor}
	a(t) = a_b \cosh\left(H_{\textrm{inf}}t\right),
\end{equation}
where $a(t)$ is the scale factor, $a_b$ is a constant quantifying the size of the universe at the bounce. The parameter $H_{\textnormal{inf}}$ is related to the energy scale of inflation just after the bounce.

In a Friedmann-Lema\^{\i}tre-Robertson-Walker (FLRW) universe with a spatially flat metric and the scale factor defined as in Eq.~\eqref{eq: scalefactor}, the minimization of the $f(R)$ action \eqref{eq: fR_action} leads to a second order differential equation for the function $f$. Solving this equation in conjunction with appropriate physical constraints gives \cite{BouhmadiLopez:2012qp}:
\begin{equation}
	\label{eq: f(R)}
	f(r) = 2H_{\textrm{inf}}^2\sqrt{r-3}
	\cos\left[\frac{\sqrt{3}}{2}\left(\pi-\arccos\frac{9-r}{3} \right)
	- \arccos\sqrt{\frac{3}{2}\frac{r-6}{r-3}}\right].
\end{equation}
In the above equation $r\equiv R/H_{\textrm{inf}}^2$.

\begin{figure*}[h]
	\centering
	\subfloat{\includegraphics[width=.42\columnwidth]{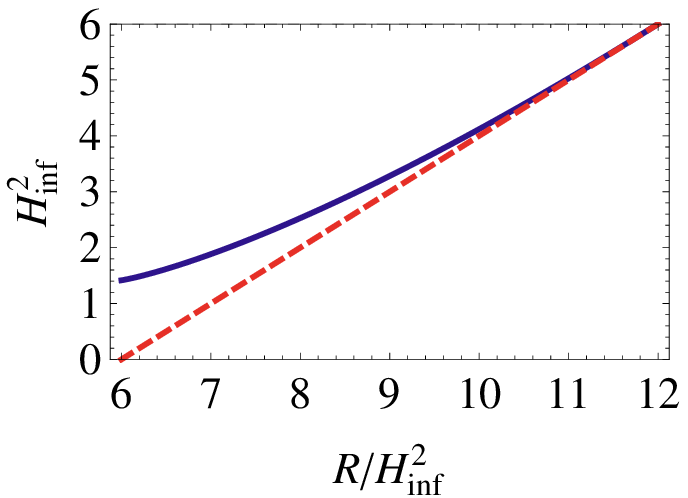}}
	\qquad
	\subfloat{\includegraphics[width=.44\columnwidth]{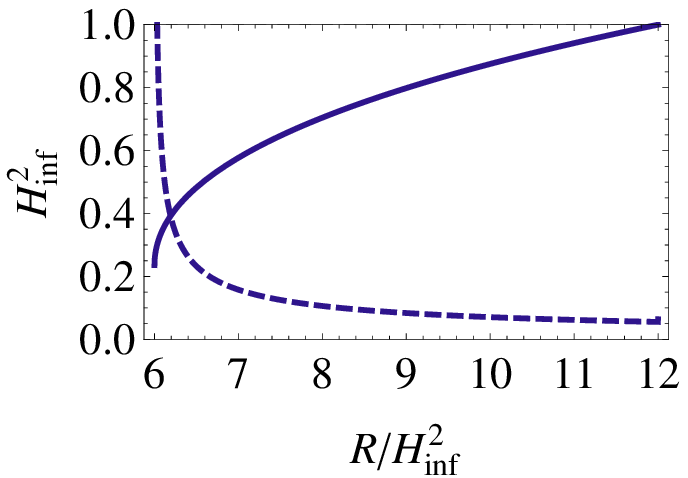}\label{fig: f_R}}
	\caption[$f(R)$]{\label{fig: f(R)}These plots show: \textbf{(left)} the behaviour of $f(R)$ as a function of $R/H_{\textnormal{inf}}^2$ (see the blue curve) and linear approximation $R-6H_{\textnormal{inf}}^2$, well during inflation (see the red curve); \textbf{(right)} the behaviour of $f_R$ (see the continuous curve) and $f_{RR}$ (see the dashed curve) as functions of $R/H_{\textnormal{inf}}^2$.}
\end{figure*}


\section{Energy spectrum of the gravitational waves}
\label{sec:2}

The spectrum of the gravitational waves is determined using the method of the continuous Bogoliubov coefficients $\alpha$ and $\beta$, as in Ref.~\cite{Parker1}. The graviton density of the universe is given by $|\beta|^2$, while the dimensionless logarithmic energy spectrum of the gravitational waves (GW) of angular frequency $\omega$ is defined at the present time $\eta_0$ as \cite{Allen:1987bk}:
\begin{equation}
\label{eq: Spectrum}
	\Omega_{GW} (\omega,\eta_0) = \frac{\hbar\kappa^2}{3\pi^2 c^5H^2(\eta_0)}\omega^4|\beta(\eta_0)|^2
\end{equation}

\begin{figure*}[t]
	\centering
	\subfloat{\includegraphics[width=.48\textwidth]{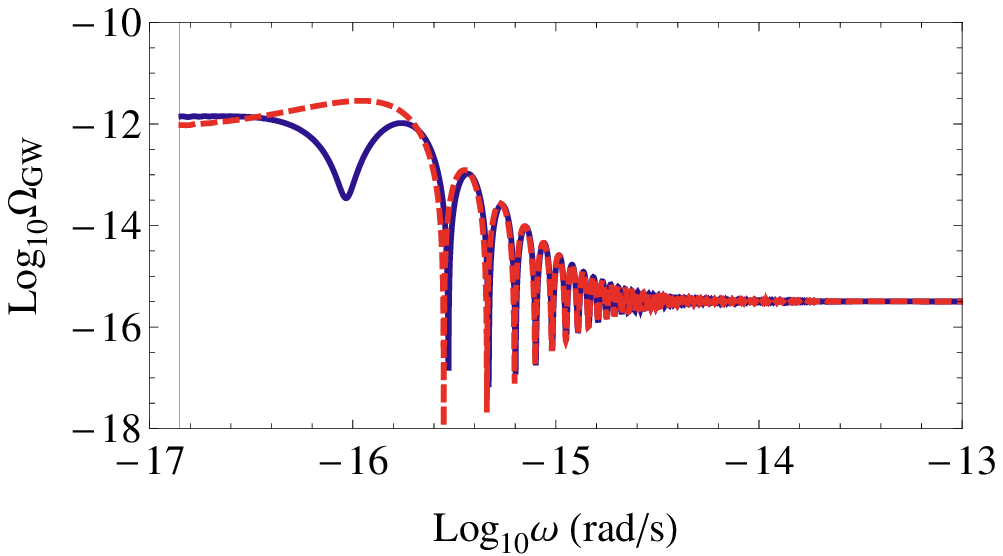}}
	\quad
	\subfloat{\includegraphics[width=.48\textwidth]{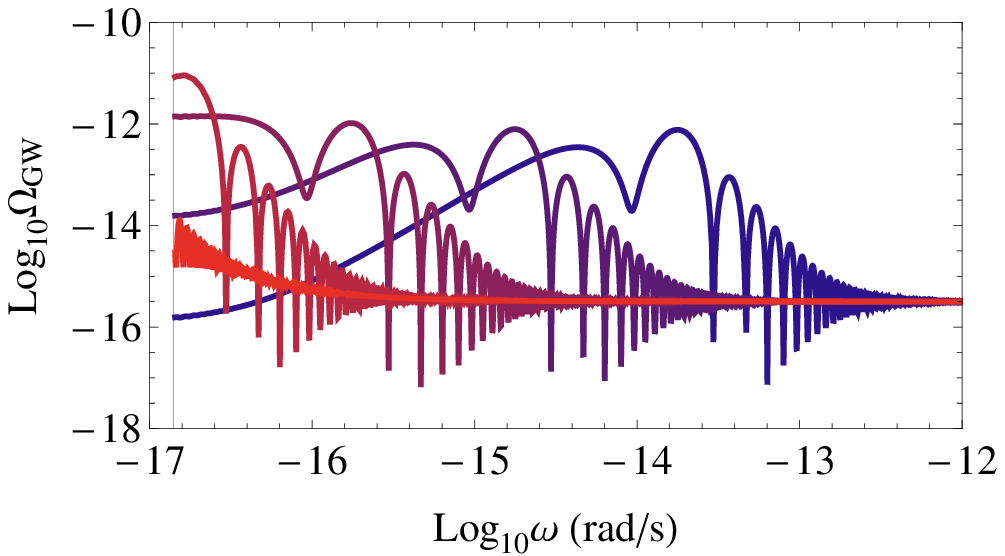}}\\
	
	\subfloat{\includegraphics[width=.48\textwidth]{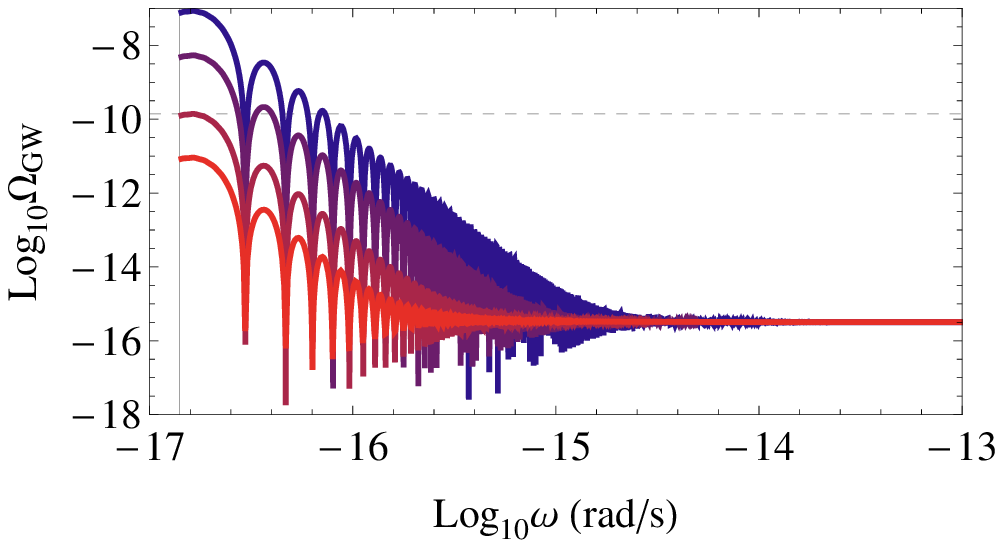}}
	\quad
	\subfloat{\includegraphics[width=.48\textwidth]{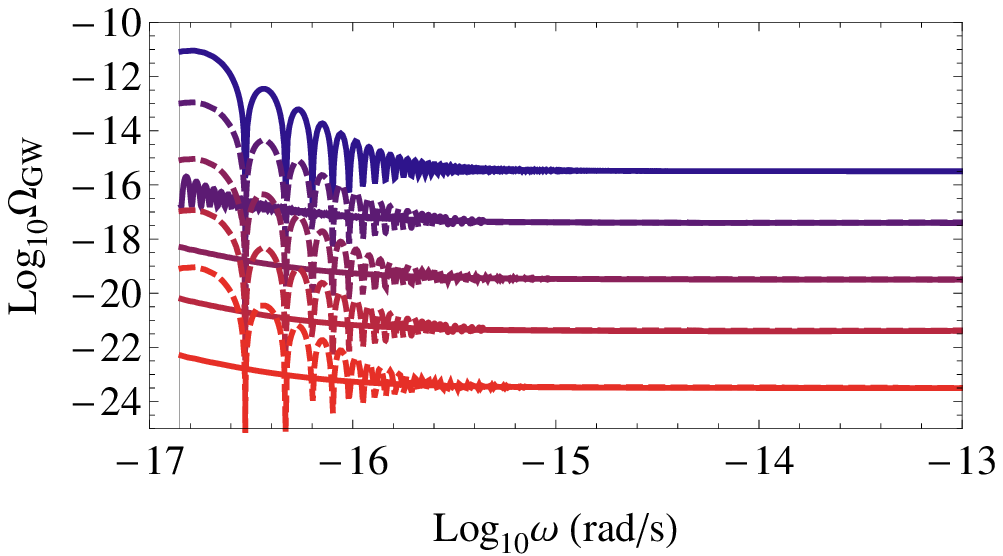}}
	\caption[Energy spectra of GW]{\label{Spectra}These plots show the spectra of the energy density of GW at present time:
	\textbf{(top~left)} Obtained with an $f(R)$ treatment  for the perturbations (blue continuous curve) and a GR treatment for the perturbations (red dashed curve). $E_{\textnormal{inf}}=1.5\times 10^{16}$GeV. $t_{\textnormal{ini}}=-\textrm{arccosh}(10)H_{\textnormal{inf}}^{-1}$. $a_b=2\times10^3$.
	\textbf{(top~right)} Otained in an $f(R)$ treatment for different values of $a_b$. The value of $a_b$  increases from the red curve to the blue curve as: $a_b=2\times 10^2$; $a_b=2\times 10^3$; $a_b=2\times 10^4$; $a_b=10\times 10^5$. $a_b=10\times 10^6$. $E_{\textnormal{inf}}=1.5\times 10^{16}$GeV. $t_{\textnormal{ini}}=-\textrm{arccosh}(10)H_{\textnormal{inf}}^{-1}$.
	\textbf{(bottom~left)} Obtained in an $f(R)$ treatment for different values of $t_{\textnormal{ini}}$. The value of $t_{\textnormal{ini}}$  increases in absolute value from the red curve to the blue curve as: $t_{\textnormal{ini}}=-\textrm{arccosh}(10)H_{\textnormal{inf}}^{-1}$; $t_{\textnormal{ini}}=-\textrm{arccosh}(20)H_{\textnormal{inf}}^{-1}$; $t_{\textnormal{ini}}=-\textrm{arccosh}(50)H_{\textnormal{inf}}^{-1}$; $t_{\textnormal{ini}}=-\textrm{arccosh}(100)H_{\textnormal{inf}}^{-1}$. $E_{\textnormal{inf}}=1.5\times 10^{16}$GeV. $a_b=2\times10^3$.
	\textbf{(bottom~right)} Obtained in an $f(R)$ treatment for different values of $E_{\textnormal{inf}}$. A comparison is made between the results obtained with a fixed value of $a_b$ (continuous curves) and a fixed value of $a_bH_{\textnormal{ini}}$ (discontinuous curves). The value of $E_{\textnormal{inf}}$  increases from the red curve to the blue curve: $E_{\textnormal{inf}}=1.5\times 10^{14}$GeV; $E_{\textnormal{inf}}=0.5\times 10^{15}$GeV; $E_{\textnormal{inf}}=1.5\times 10^{15}$GeV; $E_{\textnormal{inf}}=0.5\times 10^{16}$GeV; $E_{\textnormal{inf}}=1.5\times 10^{16}$GeV. $(a_{\textnormal{ini}}=10a_b$. $a_b=2\times 10^3)$.}
\end{figure*}

To calculate the evolution of the gravitational waves, we express the the continuous Bogoliubov coefficients in terms of the variables $X$ and $Y$, see Refs.~\cite{Sa1,Bouhmadi1}, which obey the set of  differential equations:

\begin{equation}
	\label{eq4: X''eq}
	X'' = \left(k^2 - \frac{z''}{z}\right)X,
	~~~~~~\textrm{and}~~~~~~
	X' = -ikX.
\end{equation}
Here, $k$ is the wave-number, a prime indicates a derivative with respect to the conformal time $\eta$ ($a =dt/d\eta$) and $z\equiv a\sqrt{f_R}$. The differential equations \eqref{eq4: X''eq} are integrated from an initial time $t_{\textrm{ini}}$, set before the bounce, until the present time. We describe the late time evolution of the universe in a GR setup, using the $\Lambda$CDM model \cite{WMAP} complemented with a radiation phase and making the connection between the $f(R)$ driven early inflation and the radiation phase with a model of a modified Generalized Chaplygin Gas suitable for the early universe \cite{Bouhmadi1}.
The results obtained for the GW spectra are shown in Fig.~\ref{Spectra}.

\section{Conclusions}

The existence of the bounce in the early universe affects the spectrum of GWs only in the low frequency range, $\lesssim 10^{-12}$Hz, where various peaks appear whose position and intensity depend on the parameters of the mode. The fact that the oscillatory structure appears in the spectra of (i) the GR treatment and (ii) the $f(R)$ treatment suggests it is not a consequence of the effects of $f(R)$-gravity. Similar oscillations have been obtained in works of loop quantum cosmology first pointed out by Afonso et al \cite{Afonso1}. Due to the low energy density of the cosmological GW's, the results obtained in this work are hard to be detected in the near future (see Fig. 2 of Ref.~\cite{Smith1} and Fig. 6 of Ref.~\cite{Kawasaki:2012rw}). The detection of the B-mode polarization of the CMB radiation seems to be the best candidate to obtain information about the cosmological GW's. \cite{Bourhrous:2012kr}


\begin{acknowledgement}
M.B.L. is supported by the Spanish Agency ``Consejo Superior de Investigaciones Cient\'{\i}ficas" through JAEDOC064.
This work was supported by the Portuguese Agency ``Fund\c{c}\~{a}o para a Ci\^{e}ncia e Tecnologia" through PTDC/FIS/111032/2009.
\end{acknowledgement}


\end{document}